12-30-2021

# Awareness of Predatory Journals in Library and Information Science Faculties in India

Madhuri Kumari  
*Central University of Gujarat*, madhurik909@gmail.com

Subaveerapandiyan A  
*Regional Institute of Education Mysore*, subaveerapandiyan@gmail.com

# Awareness of Predatory Journals in Library and Information Science Faculties in India

**Madhuri Kumari** and **Subaveerapandiyan A**
PhD Research Scholar, School of Library and Information Science,
Central University of Gujarat, India
Library Professional Assistant, Regional Institute of Education Mysore, India
E-mail:subaveerapandiyan@gmail.com

**Abstract**

*Predatory journals that pretended to resemble refereed journals but are used for money-making purposes. Predatory publishers produce less quality scientific and research papers; it is a severe academic threat in scientific publications. Researchers are ensuring the quality of the journal and peer-reviewing process before submitting the manuscript. This paper aims to know the Indian Library and Information Science faculties awareness and knowledge about Predatory journals. To this study, 67 LIS (Library and Information Science) faculties took part, and they are working as Assistant Professors (67.2%), Associate Professors (16.4%) and Professors (16.4%) of various states (31.3%) and central universities (68.7%) in India. The study results found 89.6% of faculty knew the term Predatory, 80.6% knew how to identify the predatory, most of them knew differing predatory, 92.5% of respondents were aware of the open access system, and most of them knew legitimate journals. The study's findings revealed LIS faculties knew predatory journals publish a high number of low-quality papers without proper peer review. The T value is -13.22, and the faculty members' opinions based on self-awareness of publishing papers have a significant highest mean value of 3.71.*

## 1. INTRODUCTION

Beall used the term predatory journals (PJs) and predatory publishers in the year 2012, published in the journal Nature. The experiment of open access has been done to freely accessible research products to the research community. As we know that every aspect has some merits and demerits, this open access comes with the unwanted practice of predatory journals. The predatory journals are imposters of the open-access model where an author has to pay article-processing charges (APCs) to publish their papers (Beall, 2012). The quality of PJs is compromised as they do not follow the ethical practices of article publishing like peer- review and follow some unethical practices like data frication, falsification, ghost authoring and plagiarism. This type of journal is known as predatory journals and publishers as predatory publishers (Hebrang Grgić & Guskić, 2019). The research community is the producer and consumer of information. If the production of false and wrong research occurs, it will be consumed in the same way, and ultimately the purpose of the study will collapse.

According to Jeffrey Beall, the predatory journals made a similar website to the legitimate journal website. The set-up of the website is so identical that it became difficult for a user to differentiate between the predatory and legitimate websites of journals. The name of the country mentioned by Beall is India, including Pakistan and Nigeria, where predatory journals are actively publishing the manuscript in the highest numbers (Beall, 2012). So, it is become imperative to find awareness among the faculty, researcher, and scholar. Scholarly article publications are compulsory for faculty members in India. Basic eligibility for Assistant professors have at least two research papers and two conference/seminar publications based on their PhD work; Associate professor minimum seven publications of peer-reviewed or UGC listed journals; Professor at least ten publications in peer-reviewed or UGC care listed (UGC, 2018). So as per UGC guidelines, research publications play a prime role in getting jobs in any academic institute/university in India. This study helped to know the situation of awareness about predatory journals among faculty of LIS in India.

Predatory publishers are used to attracting new researchers by sending fake marketing emails. They gave phoney promises to the researchers, the journal indexed in Scopus, Web of Science, PubMed, and many more. Still, the reality is not like that, so they called fraudulent journals, hijacked journals and cloned journals. Predatory journals use the unawareness of researchers knowledge of journal publications and the gap in finding out suitable journals for their publications. They underestimating young researchers journal literacy, so that still growing gigantically. Researchers have plenty of opportunities to publish a manuscript even though doubt arises in their minds on choosing suitable journals for their study. Predatory journals look legitimate, but they have many loopholes. Before submitting the manuscript, researchers have to check their journal official websites and databases. They also need to study journal guidelines and verify with indexing databases.

### 1.1 Research Question

**A)** What is the awareness and perception towards the predatory journals?

### 1.2 Hypothesis ($H_0$)

$H_0$: LIS faculties are aware of predatory journals.

### 1.3 Statement of Problem

The primary purpose of a predatory journal is to entice the authors by charging an article processing fee to publish their research articles without quality peer review, which ultimately disgrace the faculty, research, subject field and institutions (Ross-White et al., 2019). It is the concerned area to find the perception and attitude of LIS faculties towards the predatory journals. Publishing is increasing significantly in developing and low-income countries. Demir (2018) found that India had the highest number of publications in predatory journals in 2017. It is an alarming situation to check this issue. This study has been conducted to know and identify the awareness among Indian Library and Information Science faculties.

### 1.4 Aim and Objectives

Predatory journals are primarily focused on profit-making rather than peer review or quality of research. The publication is an integral part of faculty promotion and reputation. Publication in these journals may negatively impact the image or position of the faculty. So, this study aims to know the status of awareness about predatory journals among the LIS faculty of India. This study has been conducted to achieve the following objective.

**1.4.1.** To understand the perceptions and attitude towards the predatory journals and open access publishing of LIS faculty in India.

### 1.5 Scope and Limitations

The study sample limits India's library and information science faculties, working in the state and central universities. The subject domain Library and information science deal with the library professionals and researchers, faculty so, apt to know the status about awareness of predatory journals. This study is limited to the LIS faculty of India only. The research would be an integral part of any subject domain to be conducted in any subject area.

## 2. MATERIALS AND METHOD

This study is quantitative research. For this study, we used a purposive sampling technique and survey method. Samples are Indian Library Science teaching faculties from various States and Central Universities in India. A total of 16 questions was prepared and asked in this questionnaire. The structured questionnaire comprises four parts. Part one general information (gender, educational qualification, institute type and academic designation); part two scientific publishing; part three open access publishing and part four level of awareness of and about predatory journals. This research respondent is 61.2% (41) male and 38.8% (26) female. Respondents are 67.2% (45) Assistant Professors, 16.4% (11) Associate Professors and 16.4% (11) Professors. 91% of the respondents completed their PhD. We conducted an online survey from September to October 2021. We used Google Form and links, shared by faculties official and personal email IDs; for data analysis, we used frequency, percentage, two-tailed independent t-tests conducted based on their educational qualifications. Google sheet and SPSS software were used.

## 3. RESULTS AND DISCUSSION

**Table 1. Socio-Demographic Details**

| Gender | Respondents (%) |
|---|---|
| Male | 41 (61.2%) |
| Female | 26 (38.8%) |

| Highest Educational Qualifications | Respondents |
|:---:|:---:|
| PhD | 61 (91%) |
| M.Phil. | 1 (1.5%) |
| Master Degree | 5 (7.5%) |
| **Institute Type** | **Respondents** |
| Central University | 46 (68.7%) |
| State University | 21 (31.3%) |
| **Academic Designation** | **Respondents** |
| Professor | 11 (16.4%) |
| Associate Professor | 11 (16.4%) |
| Assistant Professor | 45 (67.2%) |

**Figure 1. Socio-Demographic Details**

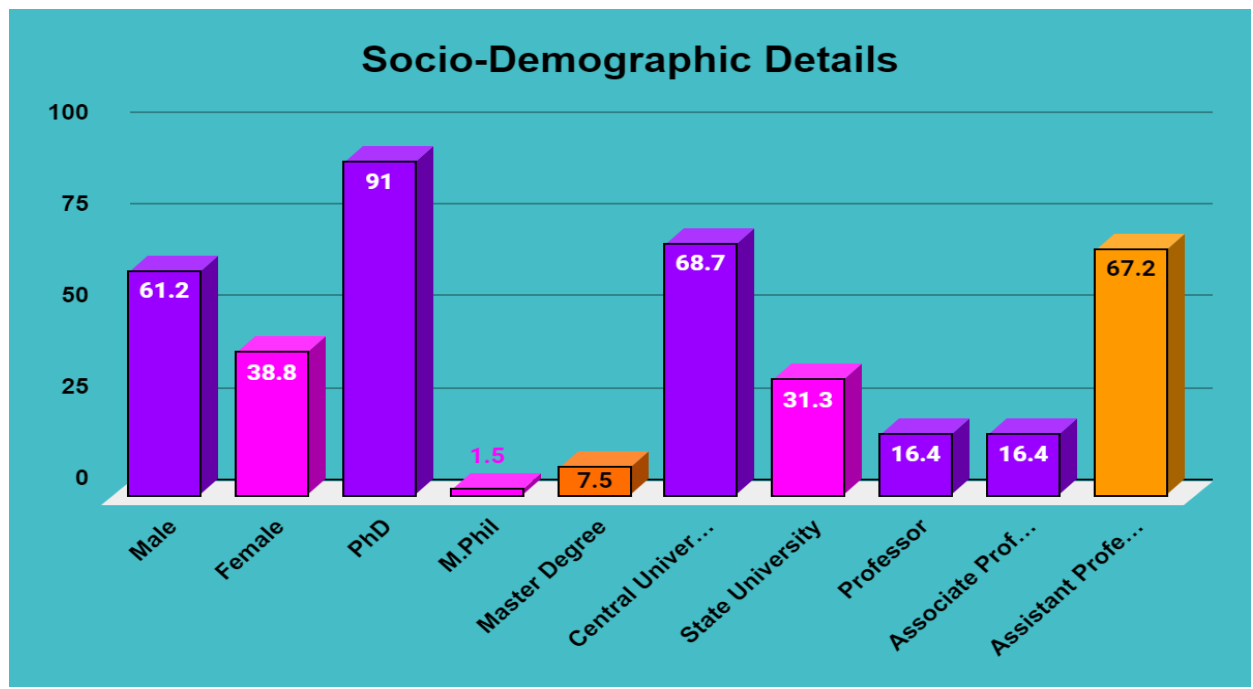

Table 1 and figure 1 presents the demographic details of the respondents. The highest number of respondents is 61.2% male, 91% of respondents completed their PhD, 68.7% of respondents working in Central Universities in India, and 67.2% of respondents are working as Assistant professor rank.

**Table 2: How many publications do you have to your credit as the first author and co-author**

| How many publications do you have to your credit as the first author and co-author? | First Author | Co-Author |
|---|---|---|
| | Respondents (%) | Respondents (%) |
| <10 publications | 26 (38.8%) | 39 (58.2%) |
| 10–20 publications | 16 (23.9%) | 14 (20.9%) |
| 21–50 publications | 16 (23.9%) | 12 (17.9%) |
| 51-100 publications | 6 (9%) | 2 (3%) |
| >100 publications | 3 (4.4%) | 0 (0%) |

Table 2 shows less than 10 publications authored as first authors (38.8%) and co-author (58.2%). The percentage of co-authors is higher than first-authored publications.

**Table 3. Which arguments are the most important for you if considering a specific journal for publication of your scientific work?**

| Most important for you to consider a specific journal for publication | Respondents (%) |
|---|---|
| Peer-review process | 26 (38.8%) |
| Impact factor of the journal | 30 (44.8%) |
| Amount of publication costs | 9 (13.4%) |
| Publishing experiences of other colleagues | 7 (10.4%) |
| Good editorial support | 15 (22.4%) |
| Good indexing (Scopus, Web of Science, PubMed, PMC, etc.) | 35 (52.2%) |
| Quality of the submission system | 14 (20.9%) |
| Scope of the journal | 21 (31.3%) |
| All of the above | 28 (41.8%) |

**Figure 2. Most important for you to consider a specific journal for publication**

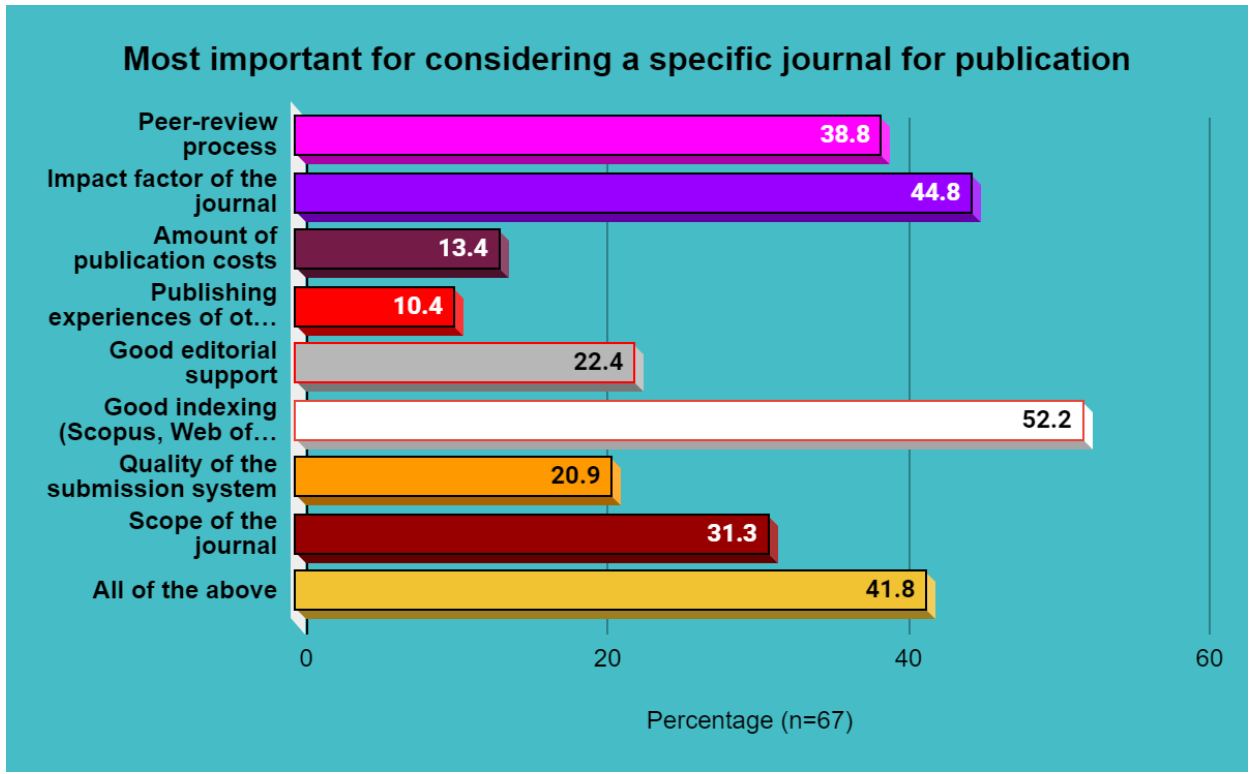

Table 3 and figure 2 shows that "Good indexing" is the most crucial point before consideration of specific journals that it is a legitimate journal by 52% of respondents. Nearly 44.8% take into consideration the impact factor of the journals before publication. Mentioning all points is important before publishing any article is considered by 41.8% of respondents.

**Table 4. Open-access publishing**

| Do you know the concept of the "Open Access" system? | Respondents (%) |
|---|---|
| Yes | 62 (92.5%) |
| No | 5 (7.5%) |
| **If yes, did you publish anything in an Open Access Journal yet?** | **Respondents (%)** |
| Yes | 47 (70.1 %) |
| No | 13 (19.4%) |
| Don't know / Not sure | 5 (7.5%) |
| Did not publish anything | 2 (3.1%) |

| How much would I be willing to pay for a publication in an Open Access Journal? | Respondents (%) |
|---|---|
| Rs Zero or nothing | 46 (68.7%) |
| <₹1000 | 5 (7.5%) |
| ₹1001-2500 | 10 (14.9%) |
| ₹2501-5000 | 3 (4.5%) |
| ₹5001-10000 | 2 (3%) |
| ₹10001-15000 | 0 (0%) |
| Above ₹15000 | 0 (0%) |
| Not Sure | 1 (1.4%) |

Above table 4 discusses open access publishing. In this, 92.5% of faculties know about the open-access system, in which 70.1% have published their papers in open access journals. The highest 68.7% of faculty don't want to pay any fee or article processing charges to publish in Open Access journals; only 29.9% are willing to spend some amount to publish their papers.

**Figure 3. Awareness about predatory journals**

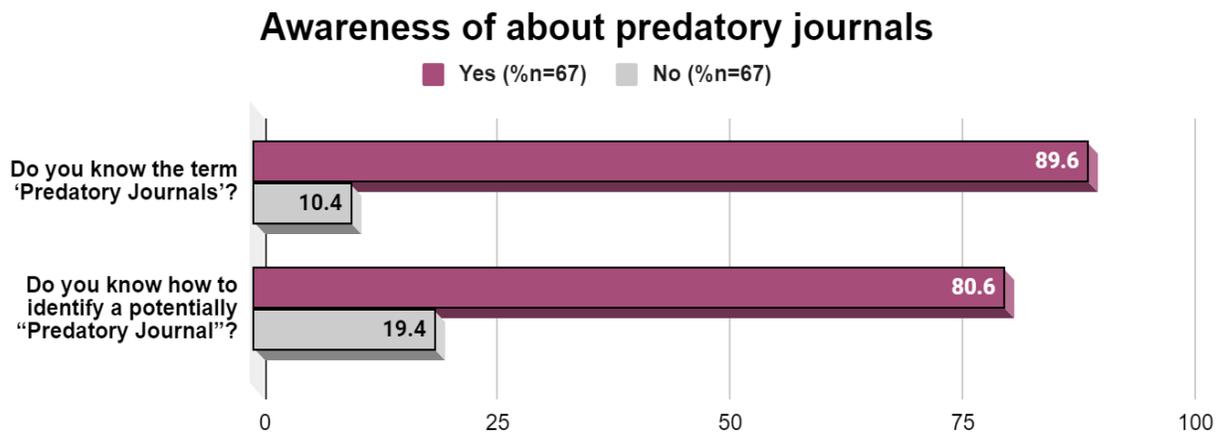

The above graph (figure 3) of awareness about predatory journals indicates that 89.6% of faculty respondents know about the term predatory journals. Still, on the question of how to identify the potential predatory journals, 80.6% are those who know about predatory journals. Still, a total of 19.4% of faculty don't know how to identify the likely predatory journals.

**Figure 4. How would you characterize a "Predatory Journal"?**

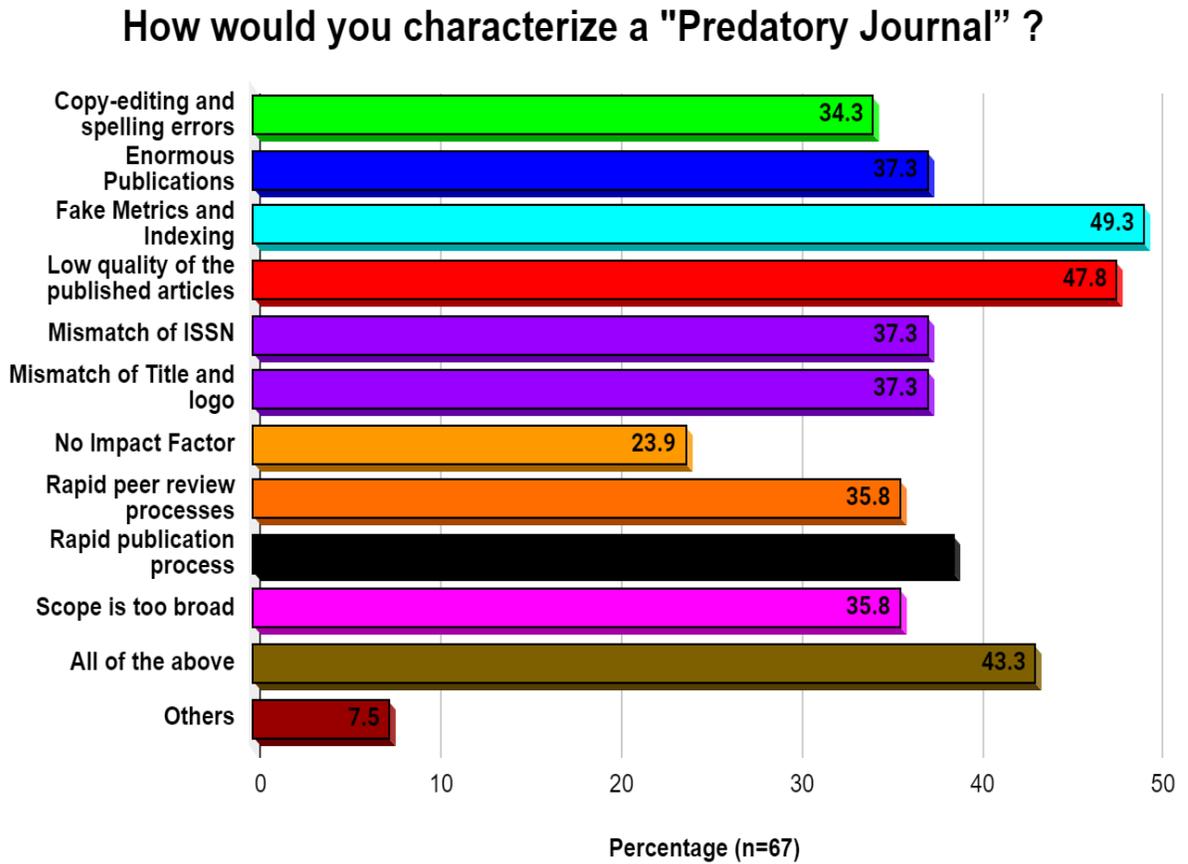

Fake metrics and indexing are mostly opted characteristic of predatory by 49.3% faculty members whereas 43.3% think that all copy-editing and spelling errors, enormous publication, low quality of published articles, mismatch of ISSN, mismatch of title and logo, no impact factor, rapid peer review, and broad scope all are the characteristics of predatory journals (figure 4).

**Table 5. Statement or criterion used for differentiating Predatory Journals (PJs) from legitimate journals**

| Statement or criterion used for differentiating Predatory Journals (PJs) from legitimate journals | Mean | SD | T-Value Qualification |
|---|---|---|---|
| I know PJs publish a high number of low-quality manuscripts and without Peer review | 3.58 | 0.84 | -13.22 |
| I know PJs' contact information is not clearly stated on the journal website | 3.76 | 1.11 | -12.61 |
| I know PJ's scope might be broad, and sometimes they may use the word international | 3.71 | 1.16 | -11.68 |
| I know PJs may add fake (non-existing) editors or the names of well-known authors without their approval | 3.64 | 1.2 | -10.65 |
| I know the peer review process may not be clearly stated on the PJ's website | 3.8 | 1.13 | -12.59 |
| I know the negative impact of PJs on my career, institution, and body of knowledge | 3.77 | 1.11 | -12.73 |
| I know PJs may use false impact factors and false location to attract manuscript submissions. | 3.74 | 1.14 | -12.02 |
| I know Beall's List Websites providing predatory journals details | 2.86 | 1.27 | -6.23 |
| I know PJs do not maintain the creative commons license | 3.35 | 1.22 | -9.09 |
| I know PJs use fake ISSN/Logo/Title or Alphabetical mismatch in titles | 3.67 | 1.21 | -10.76 |

Above table 5 shows the criteria used for differentiating predatory journals (PJs) from legitimate journals. Two-tailed independent t-tests were conducted based on the faculty's educational qualification and awareness about the differentiating Predatory Journals (PJs) from legitimate journals. The results revealed significance in each question. The highest negative t-value is -13.22. It means respondents knew Predatory journals publish a high number of low-quality manuscripts without peer review. They have less awareness about Beall's List Websites providing predatory journals details because the T-value is -6.23.

**Table 6. Give General opinion about predatory publishing, avoiding and awareness**

| Give your opinions | SA | A | DA | SD | Mean | SD |
|---|---|---|---|---|---|---|
| Conducting awareness programmes to create responsibility among the libraries | 46 (68.7%) | 16 (23.9%) | 3 (4.4%) | 2 (3%) | 3.58 | 0.84 |
| Conducting awareness programmes to create responsibility among the publishers | 43 (64.2%) | 19 (28.4%) | 3 (4.4%) | 2 (3%) | 3.53 | 0.84 |
| Conducting awareness programmes to create responsibility among the faculties | 48 (71.6%) | 16 (23.9%) | 1 (1.5%) | 2 (3%) | 3.64 | 0.81 |
| Self-Awareness is important in publishing | 50 (74.6%) | 16 (23.9%) | 0 (0%) | 1 (1.5%) | 3.71 | 0.73 |
| Reasons for the increase in Predatory journals is that some standard journals are collecting more APC | 30 (44.8%) | 28 (41.8%) | 6 (9%) | 3 (4.4%) | 3.26 | 0.89 |
| Reasons for the increase in Predatory journals is that some standard journals editors are not responding to the queries | 28 (41.8%) | 23 (34.3%) | 9 (13.4%) | 7 (10.5%) | 3.07 | 0.98 |
| Reasons for the increase in Predatory journals is that some standard journals are taking many months and years for publishing | 35 (52.3%) | 24 (35.8%) | 5 (7.5%) | 3 (4.4%) | 3.35 | 0.9 |
| Reasons for the increase in Predatory journals is that some standard journals are retaining the copyrights with them instead of giving them to the authors | 29 (43.3%) | 19 (28.3%) | 15 (22.4%) | 4 (6%) | 3.08 | 0.96 |

SA-Strongly Agree; A-Agree; DA-Disagree; SD-Strongly Disagree; SD-Standard Deviation

Table 6 shows the attitude towards general opinion about predatory publishing, avoiding and awareness. This highest mean value is 3.71, so respondents' contention is self-awareness is essential before submitting the manuscripts.

## 4. CONCLUSIONS

There have been studies on predatory OA journals like editorials and commentaries (Cobey et al., 2018) also cross-faculty awareness surveys conducted (Swanberg et al., 2020). This study is also a faculty awareness survey, but the scope and sample belong to the library and information science (LIS). The awareness among LIS faculties is getting importance due to their field of study. They educate LIS professionals as well as publish their research papers. LIS professionals directly deal with the researchers and faculty of their concerned institutions and organizations.

This study shows that LIS faculty are aware of PJs. The null hypothesis "LIS faculties are aware of predatory journals" is true because the t-value and SD are the lowest. This shows that that the LIS faculties are aware of predatory journals. They know how to identify legitimate journals before the publication of a manuscript. But India is on top in PJs publication, so it's imperative to understand the attitude among faculties. The PJs awareness programme could be a considerable step towards LIS faculty to educate about the journal publication process and quality distribution of their contribution. Awareness among LIS faculty would advocate and educate LIS students about the Predatory journals, also needed because of the prospect of future library professionals taught by these faculties. They would be equipped with, how to advocate library users like faculty and research scholars about predatory journals.

This survey did not include the faculty experience with the publishing in PJs, which could be a prospect of the survey. Authors must learn to evaluate journals based on a variety of factors, including editorial oversight (journal editors and editorial board members), peer-review practices, quality of published articles, access and indexing, metrics and citations, costs, and, most importantly, ethical practices, regardless of the publishing model (Christopher & Young, 2015). We can teach authors critical article evaluation and crucial aspects of publishing through workshops and mentorship, guiding them to avoid predatory journals and select the best ones.

# Appendix

**Part-1 Demographic characteristics (5 Questions)**

1. What is your gender identity?
a) Male b) Female
2. What is your highest qualification?
a) PhD b) M.Phil. c) Master Degree
3. Which institute are you working at?
a) Central University b) State University c) Others (Please Specify)
4. What is your present designation in your institute?
a) Professor b) Associate Professor c) Assistant Professor

**Part-2 Scientific publishing (3 Questions)**

1. How many publications do you have to your credit as the first author?
a) <10 publications b) 10–20 publications c) 21–50 publications d) 51-100 publications e) >100 publications
2. How many publications did you co-author?
a) <10 publications b) 10–20 publications c) 21–50 publications d) 51-100 publications e) >100 publications
3. Which arguments are the most important for you if considering a specific journal for publication of your scientific work?
a) Peer-review process b) Impact factor of the journal c) Amount of publication costs d) Publishing experiences of other colleagues e) Good editorial support f) Good indexing (Scopus, Web of Science, PubMed, PMC, etc.) g) Quality of the submission system h) Scope of the journal i) All of the above

## Part-3 Open-access publishing (3 Questions)

1. Do you know the concept of the "Open Access" system?
a) Yes b) No
2. If yes, have you published anything in an Open Access Journal yet?
a) Yes b) No c) Don't know / Not sure d) Did not publish anything
3. How much would I be willing to pay for a publication in an Open Access Journal?
a) Rs Zero or nothing b) <₹1000 c) ₹1001-2500 d) ₹2501-5000 e) ₹5001-10000 f) ₹10001-15000 g) Above ₹15000 h) Others

## Part-4 Awareness of about predatory journals (10 Questions)

1. Do you know the term 'Predatory Journals'?
a) Yes b) No
2. Do you know how to identify a potentially "Predatory Journal"?
a) Yes b) No
3. How would you characterize a "Predatory Journal"?
a) Copy-editing and spelling errors
b) Enormous Publications
c) Fake Metrics and Indexing
d) Low quality of the published articles
e) Mismatch of ISSN
F) Mismatch of Title and logo
g) No Impact Factor
h) Rapid peer review processes
i) Rapid publication process
j) Scope is too broad
k) All of the above
4. Statement or criterion used for differentiating Predatory Journals (PJs) from legitimate journals
i. Not at all aware ii. Slightly aware iii. Somewhat aware iv. Moderately aware v. Extremely aware)
i) I know PJs publish a high number of low-quality manuscripts and without Peer review
ii) I know PJs' contact information is not clearly stated on the journal website
iii) I know PJ's scope might be broad, and sometimes they may use the word international or global
iv) I know PJs may add fake (non-existing) editors or the names of well-known authors without their approval
v) I know the peer review process may not be clearly stated on the PJ's website
vi) I know the negative impact of PJs on my career, institution, and body of knowledge
vii) I know PJs may use false impact factors and false location to attract manuscript submissions.
viii) I know Beall's List Websites providing predatory journals details
ix) I know PJs do not maintain the creative common license
x) I know PJs use fake ISSN/Logo/Title or Alphabetical mismatch in titles
5. Give your general opinions
(Strongly Agree, Agree, Disagree, Strongly Disagree)
i) Conducting awareness programmes to create responsibility among the libraries
ii) Conducting awareness programmes to create responsibility among the publishers
iii) Conducting awareness programmes to create responsibility among the faculties

iv) Self-Awareness is important in publishing
v) Reasons for the increase in Predatory journals is that some standard journals are collecting more APC
vi) Reasons for the increase in Predatory journals is that some standard journals editors are not responding to the queries
vii) Reasons for the increase in Predatory journals is that some standard journals are taking many months and years for publishing
viii) Reasons for the increase in Predatory journals is that some standard journals are retaining the copyrights with them instead of giving them to the authors
6. Any other comments/suggestions/opinions (Optional)